\documentstyle[12pt,a41,epsfig,wrapfig,rotating]{article}

\begin{document}

\begin{center}
{\large \bf Target Mass Corrections to the Spin-dependent Structure
Functions}\footnote{Work supported in part by EU 
contract FMRX-CT98-0194(DG 12 - HIHT)}
\\
\vspace*{3mm}
{\bf J.~Bl\"umlein and A.~Tkabladze\footnote{
Alexander von Humboldt Fellow}}\\
\vspace*{3mm}
{\it DESY Zeuthen, D-15738 Zeuthen, Germany}\\
\vspace*{3mm}
\end{center}

\renewcommand{\thefootnote}{\fnsymbol{footnote}}

At the energies at which most data in polarized deep-inelastic 
scattering have been taken  the ratio $M^2/Q^2$,
where $M$ is the mass of the nucleon, is not negligible.
The target mass corrections should be taken into account to analyze 
the $Q^2$ dependence of structure functions at low $Q^2$ and to 
extract information of dynamical higher twist effects at moderate energies.

We calculated the target mass corrections for  spin dependent structure
functions (SF)  both for neutral and charged current interactions using the 
operator product expansion in  lowest order in QCD. 
In this paper  we present only the twist-2 contribution
to the structure functions\footnote{The details, also for the operators 
of twist-3, are given in Ref. \cite{BT}.}.
We considered the light-cone expansion of the forward Compton amplitude in
 momentum space. Using the corresponding dispersion relations \cite{BK}
the moments of structure functions can be 
expressed by the xmatrix elements
of twist-2 operators.
The target mass effects are calculated using the method proposed by Georgi 
and Politzer \cite{GP}. 
 Following this way we expressed the moments of 
the SF as series of $(M^2/Q^2)^j$ and
functions of reduced operator matrix elements. 
 After performing the inverse
Mellin transform we get the following expressions for the SF:
\begin{eqnarray}
g_1^{\pm}(x)& = &
x\frac{d}{ dx} x\frac{ d}{d x}
\left[\frac{x}{(1+4M^2 x^2/Q^2)^{1/2}} \frac{G^{\pm}_1(\xi)}{\xi}\right],
\label{g1G}\\
g_2^{\pm}(x)& = &-
x\frac{d^2}{ dx^2}x
\left[\frac{x}{(1+4M^2 x^2/Q^2)^{1/2}} \frac{G^{\pm}_1(\xi)}{\xi}\right],
\label{g2G}\\
g_3^{\pm}(x) & = & 
2 x^2\frac{d^2}{ dx^2}
\left[\frac{x^2}{(1+4M^2 x^2/Q^2)^{1/2}} \frac{G^{\pm}_2(\xi)}{\xi^2}\right],
\label{g3G}\\
g_4^{\pm}(x) & = & -
x^2\frac{d}{ dx}x\frac{d^2}{dx^2}
\left[\frac{x^2}{(1+4M^2 x^2/Q^2)^{1/2}} \frac{G^{\pm}_2(\xi)}{\xi^2}\right],
\label{g4G}\\
g_5^{\pm}(x) & = & -
x\frac{d}{ dx}
\left[\frac{x}{(1+4M^2 x^2/Q^2)^{1/2}} \frac{G^{\pm}_3(\xi)}{\xi}\right]
+\frac{M^2}{Q^2}x^2\frac{d^2}{ dx^2}
\left[\frac{x^2}{(1+4M^2 x^2/Q^2)^{1/2}} \frac{G^{\pm}_2(\xi)}{\xi^2}\right].
\label{g5G}
\end{eqnarray}
The functions $G^{\pm}_1(y)$, $G^{\pm}_2(y)$ and $G^{\pm}_3(y)$ are defined 
via the matrix elements of the twist-2 operators appearing in the light-cone
expansion of the forward scattering amplitude.

The expressions for the SF $g_1(x,Q^2)$ and 
$g_2(x,Q^2)$, Eqs. (\ref{g1G}) and (\ref{g2G}), can be rewritten 
as
\begin{eqnarray}
g_1(x)&=&x\frac{d}{dx}{\cal F}(x)+x^2\frac{d^2}{dx^2}{\cal F}(x),
\label{WWg1}\\ 
g_2(x)&=&-2x\frac{d}{dx}{\cal F}(x)-x^2\frac{d^2}{dx^2}{\cal F}(x).
\label{WWg2}
\end{eqnarray}
Here  ${\cal F}(x)$ denotes the function in the brackets of  Eqs.
 (\ref{g1G}) and (\ref{g2G}).
>From  (\ref{WWg1}) and (\ref{WWg2}) one obtains
\begin{eqnarray}
g_2(x)&=&-g_1(x)-x\frac{d}{dx}{\cal F}(x). \label{WWaux1}
\end{eqnarray}
Integrating the Eq. (\ref{WWg1}) we get
\begin{eqnarray}
x\frac{d}{dx}{\cal F}(x)&= &-\int_{x}^{1}{\frac{dy}{y}g_1(y)}.
\label{WWaux2}
\end{eqnarray}
Eqs. (\ref{WWaux1}) and (\ref{WWaux2}) lead to the Wandzura-Wilczek
relation~\cite{WW}
\begin{eqnarray}
g_2(x)&= &-g_1(x)+\int_{x}^{1}{\frac{dy}{y}g_1(y)}.
\label{WW}
\end{eqnarray}
Therefore, the Wandzura-Wilczek relation is not affected by
target mass corrections. This  result was obtained in Ref. \cite{PR} before.
Moreover, it was shown that the Wandzura-Wilczek relation is valid for 
the quarkonic operators even 
in the massive quark case \cite{BT}.
Although the expression for $g_2(x,Q^2)$, Eq. (\ref{WWg2}), is formally 
consistent with the Burkhardt-Cottingham sum rule \cite{BC}
\begin{eqnarray}
\label{e155}
\int_0^1dxg^i_2(x,Q^2)=0,
\end{eqnarray}
the $0th$ moment of the structure functions $g_2^i(x,Q^2)$ is not
described by the local operator product expansion.

The relation between the structure functions $g_3(x)$ and $g_4(x)$ 
obtained by Bl\"umlein and Kochelev \cite{BK}
is also valid at any order in $M^2/Q^2$. Integrating  Eq. 
(\ref{g4G}) one gets
\begin{eqnarray}
\label{e300}
{g_3}^i(x,Q^2)=2x\int_x^1\frac{dy}{y^2}{g_4}^i(y,Q^2).
\end{eqnarray}

Unlike the above relations, the Dicus relation between
$g_4(x)$ and $g_5(x)$, cf. Ref. \cite{BK},
is violated by the target mass
corrections. Note that the Dicus relation corresponds to the 
Callan-Gross  \cite{CG}
relation for unpolarized structure functions. The spin  
 dependence enters in front of the corresponding structure functions as  
 an overall
factor $S\cdot q$. It is worth mentioning that the Callan-Gross
 relation is also
violated by target mass corrections \cite{GP}.

A.T. acknowledges  the  support  by the Alexander von Humboldt Foundation.

\end{document}